\documentclass{article}
\usepackage{waspaa17,amsmath,graphicx,url,times}
\usepackage{color}
\usepackage{bm}
\usepackage{acronym}
\usepackage{siunitx}
\usepackage[tight]{subfigure}
\usepackage{pgfplots}
\pgfplotsset{compat=newest}
\usetikzlibrary{patterns}
\usepackage{cite}

\pgfplotsset{
	colormap/parula/.style={colormap={parula}{
		rgb=(0.2081,0.1663,0.5292)
		rgb=(0.2116,0.1898,0.5777)
		rgb=(0.2123,0.2138,0.627)
		rgb=(0.2081,0.2386,0.6771)
		rgb=(0.1959,0.2645,0.7279)
		rgb=(0.1707,0.2919,0.7792)
		rgb=(0.1253,0.3242,0.8303)
		rgb=(0.0591,0.3598,0.8683)
		rgb=(0.0117,0.3875,0.882)
		rgb=(0.006,0.4086,0.8828)
		rgb=(0.0165,0.4266,0.8786)
		rgb=(0.0329,0.443,0.872)
		rgb=(0.0498,0.4586,0.8641)
		rgb=(0.0629,0.4737,0.8554)
		rgb=(0.0723,0.4887,0.8467)
		rgb=(0.0779,0.504,0.8384)
		rgb=(0.0793,0.52,0.8312)
		rgb=(0.0749,0.5375,0.8263)
		rgb=(0.0641,0.557,0.824)
		rgb=(0.0488,0.5772,0.8228)
		rgb=(0.0343,0.5966,0.8199)
		rgb=(0.0265,0.6137,0.8135)
		rgb=(0.0239,0.6287,0.8038)
		rgb=(0.0231,0.6418,0.7913)
		rgb=(0.0228,0.6535,0.7768)
		rgb=(0.0267,0.6642,0.7607)
		rgb=(0.0384,0.6743,0.7436)
		rgb=(0.059,0.6838,0.7254)
		rgb=(0.0843,0.6928,0.7062)
		rgb=(0.1133,0.7015,0.6859)
		rgb=(0.1453,0.7098,0.6646)
		rgb=(0.1801,0.7177,0.6424)
		rgb=(0.2178,0.725,0.6193)
		rgb=(0.2586,0.7317,0.5954)
		rgb=(0.3022,0.7376,0.5712)
		rgb=(0.3482,0.7424,0.5473)
		rgb=(0.3953,0.7459,0.5244)
		rgb=(0.442,0.7481,0.5033)
		rgb=(0.4871,0.7491,0.484)
		rgb=(0.53,0.7491,0.4661)
		rgb=(0.5709,0.7485,0.4494)
		rgb=(0.6099,0.7473,0.4337)
		rgb=(0.6473,0.7456,0.4188)
		rgb=(0.6834,0.7435,0.4044)
		rgb=(0.7184,0.7411,0.3905)
		rgb=(0.7525,0.7384,0.3768)
		rgb=(0.7858,0.7356,0.3633)
		rgb=(0.8185,0.7327,0.3498)
		rgb=(0.8507,0.7299,0.336)
		rgb=(0.8824,0.7274,0.3217)
		rgb=(0.9139,0.7258,0.3063)
		rgb=(0.945,0.7261,0.2886)
		rgb=(0.9739,0.7314,0.2666)
		rgb=(0.9938,0.7455,0.2403)
		rgb=(0.999,0.7653,0.2164)
		rgb=(0.9955,0.7861,0.1967)
		rgb=(0.988,0.8066,0.1794)
		rgb=(0.9789,0.8271,0.1633)
		rgb=(0.9697,0.8481,0.1475)
		rgb=(0.9626,0.8705,0.1309)
		rgb=(0.9589,0.8949,0.1132)
		rgb=(0.9598,0.9218,0.0948)
		rgb=(0.9661,0.9514,0.0755)
		rgb=(0.9763,0.9831,0.0538)
	}}
}

\acrodef{DI}{Directivity Index}
\acrodef{DoA}{Direction of Arrival}
\acrodef{DTFT}{Discrete-Time Fourier Transform}
\acrodef{EARS}{Embodied Audition for RobotS}
\acrodef{FIR}{Finite Impulse Response}
\acrodef{FSU}{Filter-and-Sum Unit}
\acrodef{fwSegSNR}{frequency-weighted segmental Signal-to-Noise Ratio}
\acrodef{HRIR}{Head-Related Impulse Response}
\acrodef{HRRIR}{Head-Related Room Impulse Response}
\acrodef{HRTF}{Head-Related Transfer Function}
\acrodef{LS}{Least-Squares}
\acrodef{MSE}{Mean Squared Error}
\acrodef{PLD}{Prototype Look Direction}
\acrodef{PFSB}{Polynomial Filter-and-Sum Beamformer}
\acrodef{PPF}{Polynomial Postfilter}
\acrodef{RIR}{Room Impulse Response}
\acrodef{RLSFI}{Robust Least-Squares Frequency-Invariant}
\acrodef{RLSFIP}{Robust Least-Squares Frequency-Invariant Polynomial}
\acrodef{1D-RLSFIP}{one-dimensional Robust Least-Squares Frequency-Invariant Polynomial}
\acrodef{2D-RLSFIP}{two-dimensional Robust Least-Squares Frequency-Invariant Polynomial}
\acrodef{SNR}{Signal-to-Noise Ratio}
\acrodef{WNG}{White Noise Gain}

\DeclareMathOperator*{\argmin}{argmin}

\definecolor{LMSdarkblue}{rgb}{0.216,0.255,1}
\definecolor{myBlue}{rgb}{0.2081,0.1663,0.5292}
\definecolor{myGreen}{RGB}{71,248,96}
\definecolor{myOrange}{RGB}{247,144,37}

\title{Design of robust two-dimensional polynomial beamformers as a convex optimization problem with application to robot audition}

\name{Hendrik Barfuss, Markus Bachmann, Michael Buerger, Martin Schneider$^1$\thanks{$^1$ M. Schneider was with the Friedrich-Alexander University Erlangen-N\"urnberg while this work was started. He is now with Fraunhofer IIS, Erlangen, Germany, as part of the International Audio Laboratories Erlangen.}, and Walter Kellermann}

\address{Multimedia Communications and Signal Processing,\\
	Friedrich-Alexander University Erlangen-N\"urnberg\\
	Cauerstr. 7, 91058 Erlangen, Germany \\
	{\ninept \{hendrik.barfuss, markus.m.bachmann, michael.buerger, walter.kellermann\}@FAU.de, martin.schneider@audiolabs-erlangen.de}}
\begin{document}

\ninept
\maketitle

%\vspace*{-12.5mm}

\begin{sloppy}

\begin{abstract}	\vspace{-1.5mm}
	%What we do
	We propose a robust two-dimensional polynomial beamformer design method, formulated as a convex optimization problem, which allows for flexible steering of a previously proposed data-independent robust beamformer in both azimuth and elevation direction.~As an exemplary application, the proposed two-dimensional polynomial beamformer design is applied to a twelve-element microphone array, integrated into the head of a humanoid robot. To account for the effects of the robot's head on the sound field, measured head-related transfer functions are integrated into the optimization problem as steering vectors.
	The two-dimensional polynomial beamformer design is evaluated using signal-independent and signal-dependent measures. The results confirm that the proposed polynomial beamformer design approximates the original fixed beamformer design very accurately, which makes it an attractive approach for robust real-time data-independent beamforming.
	\vspace{-1mm}
\end{abstract}

\begin{keywords}
	Robust superdirective beamforming, polynomial beamforming, white noise gain, robot audition
\end{keywords}

% reset acronyms
%------------------------------------------------------------------------
\acresetall

%------------------------------------------------------------------------
%Introduction
%------------------------------------------------------------------------
\section{Introduction}
\label{sec:introduction}
\vspace*{-2mm}
%introduction beamforming
When a mixture of target and interfering sources, impinging from different \acp{DoA}, is recorded by a microphone array, beamforming is an effective means to enhance the noisy target signal and suppress interference \cite{vanveen:assp1988}. When designing beamformers, it is important to control their robustness against small random errors like microphone mismatch or position errors of microphones to guarantee an acceptable signal enhancement performance in practical realizations \cite{bitzer:2001superdirective,VanTrees:book2004}. 
Among the various methods to increase robustness of beamformers, e.g., \cite{cox:aasp1987,carlson:aes1988,bitzer:2001superdirective,doclo:tsp2003,doclo:tasl2007}, constraining the beamformer's \ac{WNG} has recently become very popular, since convex optimization techniques allow for a direct integration of the \ac{WNG} constraint into the beamformer design, see, e.g., \cite{yan:jasa2007,mabande:icassp2009,crocco:eusipco2010,crocco:tsp2011,sun:tasl2011}. 

%Edwins work
In our previous work, a data-independent \ac{RLSFI} beamformer design, formulated as a convex optimization problem, has been presented in \cite{mabande:icassp2009}. The design criterion is to approximate a desired beamformer response subject to a distortionless response constraint on the target look direction and a user-defined lower bound on the \ac{WNG}, which gives the user direct control over the beamformer's robustness. Since the work was presented for linear arrays, a one-dimensional desired response, defined in a plane corresponding to a fixed elevation angle, was chosen for the design. Based on the work in \cite{kajala:icassp2001}, the \ac{RLSFI} beamformer design of \cite{mabande:icassp2009} was extended to the concept of polynomial beamforming in \cite{mabande:iwaenc2010}, yielding the \ac{1D-RLSFIP} beamformer design which allows for flexible beam steering in either azimuth or elevation direction. Further work by other groups on one-dimensional polynomial beamforming can be found in \cite{lai:iwaenc2010,lai:icassp2011,wang:ispcc2013,wang:eurasip2015}.
%application to robot audition
Recently, we applied both the \ac{RLSFI} and \ac{1D-RLSFIP} designs to a microphone array which was attached to the head of a humanoid robot \cite{barfuss:waspaa2015,barfuss:iwaenc2016}. Measured \acp{HRTF} were integrated into the respective convex optimization problem to account for the scattering effects of the robot's head on the sound field.
%1D -> 2D
Finally, the \ac{HRTF}-based (non-polynomial) \ac{RLSFI} design of \cite{barfuss:waspaa2015} was extended to two dimensions in \cite{barfuss:hscma2017}, where the desired beamformer response was defined for both azimuth and elevation angles. As a result the beamformer response can now be controlled for all \acp{DoA} on a sphere surrounding the three-dimensional microphone array integrated into the humanoid robot's head.

%contribution of this paper
In this work we extend the concept of the \ac{1D-RLSFIP} beamformer design, allowing for flexible beam steering in either azimuth or elevation direction, of \cite{mabande:iwaenc2010}, to a \ac{2D-RLSFIP} beamformer design, which provides flexible beam steering in both azimuth and elevation direction in real time. To this end, we make use of the extended non-polynomial \ac{RLSFI} beamformer design of \cite{barfuss:hscma2017}. The proposed \ac{2D-RLSFIP} beamformer design is applied to a twelve-element microphone robot head array and evaluated in a robot-audition scenario. Hence, the presented work can be seen as an extension of the combined work of \cite{mabande:iwaenc2010} (\ac{1D-RLSFIP} beamforming) and \cite{barfuss:hscma2017} (\ac{HRTF}-based non-polynomial \ac{RLSFI} beamformer design with two-dimensional beamformer response for robot audition).

%structure of paper
The remainder of this work is structured as follows: In Section~\ref{sec:robust_2Dpbf}, we present the \ac{2D-RLSFIP} beamformer design which allows for flexible beam steering in both azimuth and elevation direction. This design method is evaluated in Section~\ref{sec:evaluation}. Finally, the work is summarized and an outlook is given in Section~\ref{sec:conclusion}. \vspace*{-2mm}

%------------------------------------------------------------------------
%2D polynomial beamforming 
%------------------------------------------------------------------------
\section{Robust polynomial beamforming in two dimensions}
\label{sec:robust_2Dpbf}
\vspace*{-1.5mm}

%Concept 2D polynomial beamforming 
%------------------------------------------------------------------------
\subsection{Concept of two-dimensional polynomial beamforming}
\vspace*{-1.5mm}
\label{subsec:concept_2Dpbf}
%Beschreibung Struktur 2D-PFSB
In Fig.~\ref{fig:PFSB}, the block diagram of a two-dimensional \ac{PFSB} is illustrated. It consists of $R+1$ parallel blocks comprising $P+1$ \acp{FSU} and one \ac{PPF} each, followed by an outer \ac{PPF}. Each \ac{FSU} contains $N$ \ac{FIR} filters $\bm{w}_{n,p,r} = [w_{npr,0}, w_{npr,1}, \ldots, w_{npr,L-1}]^\mathrm{T}$ of length $L$, where $(\cdot)^ \mathrm{T}$ represents the transpose of a vector or matrix. The total number of \acp{FSU} equals $(P+1)(R+1)$, and the total number of \ac{FIR} filters is $N(P+1)(R+1)$. 
%Output signal FSUs
The output signal $y_{p,r}[k]$ of the $p,r$-th \ac{FSU} at time instant $k$ is obtained by convolving the $N$ microphone signals with the \ac{FIR} filters $\bm{w}_{n,p,r}, \, n \in \{1, \ldots, N\}$, followed by a summation over all channels.
\begin{figure}
	\centering
	\footnotesize
	\def\svgwidth{9.7cm}
	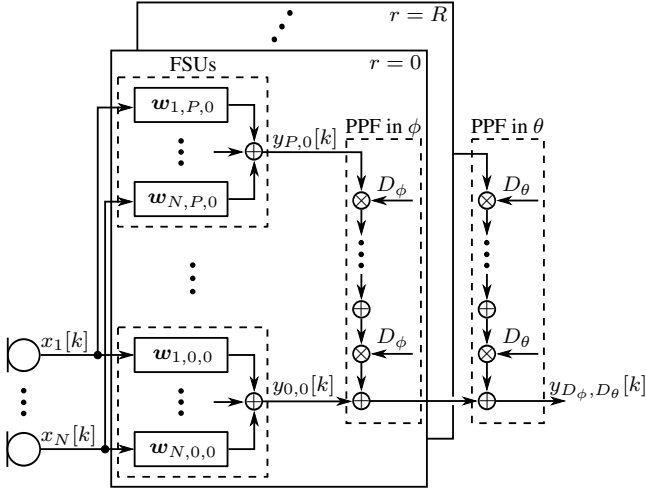 \vspace{-6pt}
	\vspace*{-4mm}
	\caption{Block diagram of a two-dimensional polynomial filter-and-sum beamformer.}
	\label{fig:PFSB}  
	\vspace*{-4.5mm}
\end{figure}
The output signal of the $r$-th block of \acp{FSU} and \ac{PPF} is given as a polynomial of order $P$ with real-valued variable $D_\phi$, where the output signals $y_{p,r}[k]$ act as coefficients of the polynomial. Analogously, these $R+1$ output signals can then be interpreted as the coefficients of an order-$R$ polynomial with real-valued variable $D_\theta$, which yields the final output signal $y_{D_\phi,D_\theta}[k]$ of the two-dimensional \ac{PFSB}:\vspace*{-1.5mm}
\begin{equation}
	y_{D_\phi,D_\theta}[k] = \sum\limits_{r=0}^{R} D_\theta^{r} \sum\limits_{p=0}^{P} D_\phi^{p} y_{p,r}[k].
	\label{eq:y_PFSB}
	\vspace*{-1.5mm}
\end{equation}
The two-dimensional \ac{PFSB} can be interpreted as follows: Each of the $R+1$ blocks of \acp{FSU} and \ac{PPF} performs one-dimensional polynomial beamforming in azimuth direction, as described in \cite{kajala:icassp2001,mabande:iwaenc2010}, for a fixed elevation angle. The $R+1$ azimuth-interpolated output signals are then interpolated in elevation direction by the outer \ac{PPF}, yielding the final output signal.
The advantage of a \ac{PFSB} is that the \ac{FIR} filters $\bm{w}_{n,p,r}$ can be designed beforehand using more complicated design methods which cannot be solved in real time, and remain fixed during runtime. The actual steering of the main beam is accomplished by simply changing the interpolation factors $D_\phi$ and $D_\theta$, which control the interpolation in azimuth and elevation direction, respectively. More details on how the interpolation factors are chosen and how the \ac{FIR} filters are designed are given in subsection~\ref{subsec:2DRLSFIP}. 
The beamformer response of the two-dimensional \ac{PFSB} is given as\vspace*{-1.5mm}
\begin{equation}
B_{D_\phi,D_\theta}(\omega,\Omega) = \sum\limits_{r=0}^{R} D_\theta^{r} \sum\limits_{p=0}^{P} D_\phi^{p} \sum\limits_{n=1}^{N} W_{n,p,r}(\omega) g_{n}(\omega, \Omega),
\label{eq:B_PFSB}
\vspace*{-1.5mm}
\end{equation}	
where $W_{n,p,r}(\omega)=\sum_{l=0}^{L-1} w_{npr,l} e^{-\mathrm{j} \omega l}$ represents the \ac{DTFT} transform of $\bm{w}_{n,p,r}$, $\mathrm{j}^2=-1$ is the imaginary unit, and $g_{n}(\omega, \Omega)$ is the sensor response of the $n$-th microphone with respect to a plane wave with \ac{DoA} $\Omega \equiv (\phi,\theta)$. Azimuth and elevation angle $\phi$ and $\theta$ are measured relative to the positive $x$- and positive $z$-axis, respectively, as in \cite{VanTrees:book2004}.
\vspace*{-2mm}

%OP
%------------------------------------------------------------------------
\subsection{Two-dimensional \acs{RLSFIP} beamformer design}
\label{subsec:2DRLSFIP}
As for the one-dimensional polynomial beamformer design in \cite{mabande:iwaenc2010}, the main goal of the proposed \ac{2D-RLSFIP} beamformer design is to approximate the non-polynomial \ac{RLSFI} beamformer design as accurately as possible for a desired angular range, while offering flexible beam steering at the same time. To this end, the design criterion is to jointly approximate $I$ desired beamformer responses $\hat{B}_{D_{\phi_i} D_{\theta_i}}(\omega, \Omega)$, each corresponding to a different \ac{PLD} $\Omega_i=(\phi_i,\theta_i), \, i \in \{1, \ldots, I\}$, by the actual beamformer response in the \ac{LS} sense. Analogously to \cite{kajala:icassp2001,mabande:iwaenc2010}, the interpolation factors are chosen as $D_{\phi_i}=(\phi_i-\ang{90})/\ang{90}$ and $D_{\theta_i}=(\theta_i-\ang{90})/\ang{90}$. Hence, $D_{\phi_i}$ and $D_{\theta_i}$ lie in the interval between $-1$ and $1$ for \acp{PLD} lying in the frontal hemisphere defined by $\ang{0} \leq \Omega \leq \ang{180}$. For interpolation parameters which do not correspond to those of the $I$ \acp{PLD}, the two \acp{PPF} will interpolate between them. For example, $(D_{\phi},D_{\theta})=(1/3,0)$ will steer the main beam towards $\Omega = (\ang{120},\ang{90})$. As a consequence, the \acp{PLD} have to be distributed across the entire angular region of interest in order to obtain an acceptable beamforming performance in this region.

In addition to the \ac{LS} approximation of the desired beamformer response, a lower bound on the \ac{WNG} and a distortionless response constraint are imposed on each of the $I$ \acp{PLD}. To obtain a numerical solution, the approximation is carried out for $Q$ frequencies $\omega_q, \, q \in \{1, \ldots, Q\}$, and $M$ design look directions $\Omega_m, \, m \in \{1, \ldots, M\}$, for which the desired beamformer response for the $i$-th \ac{PLD} is specified.
Thus, the optimization problem at frequency $\omega_q$ to be minimized can be expressed as\vspace*{-2.5mm}
\begin{equation}
	\argmin\limits_{\bm{w}_\text{f}(\omega_{q})} \sum\limits_{i=1}^{I} \Vert \bm{G}(\omega_{q}) \bm{D}_{i} \bm{w}_\text{f}(\omega_{q}) - \bm{b}_{\text{des},i} \Vert_{2}^{2},
	\label{eq:OP_LSApproximation}
	\vspace*{-1.5mm}
\end{equation}
subject to constraints on the \ac{WNG} and the beamformer response for all $I$ \acp{PLD}:\vspace*{-1.5mm}
\begin{align}
	\frac{ |\bm{a}^\text{T}_{i}(\omega_{q}) \bm{D}_{i} \bm{w}_\text{f}(\omega_{q})|^{2}}{\Vert \bm{D}_{i}\bm{w}_\text{f}(\omega_{q}) \Vert_{2}^{2}} & \ge \gamma > 0, \quad  \bm{a}^\text{T}_{i}(\omega_{q}) \bm{D}_{i} \bm{w}_\text{f}(\omega_{q}) = 1,\nonumber\\
	& \forall i = 1, \ldots, I.	\vspace*{-4mm}
	\label{eq:OP_Constraints}
\end{align}%
Eq.~(\ref{eq:OP_LSApproximation}) describes the \ac{LS} approximation of the desired beamformer response by the actual beamformer response for all $I$ \acp{PLD}, and (\ref{eq:OP_Constraints}) represents \ac{WNG} (left term\footnote{Note that the numerator of the \ac{WNG} could also be set to one due to the distortionless response constraint.}) and distortionless response constraint (right term) for each \ac{PLD}. In (\ref{eq:OP_LSApproximation}) and (\ref{eq:OP_Constraints}), the $M \times N$ matrix with $[\bm{G}(\omega_q)]_{mn}=g_n(\omega_q,\Omega_m)$ contains the sensor responses for the $M$ design look directions and $N$ microphones, vector $\bm{w}_\text{f}(\omega_q)$ of length $N(P+1)(R+1)$ contains all frequency-domain filter coefficients $W_{npr}(\omega_q)$, vector $\bm{b}_{\text{des},i}=[\hat{B}_{D_{\phi_i}D_{\theta_i}}(\omega_q,\Omega_1), \ldots, \hat{B}_{D_{\phi_i}D_{\theta_i}}(\omega_q,\Omega_M)]^\mathrm{T}$ of length $M$ includes the desired response for the $i$-th \ac{PLD}, and $\Vert \cdot \Vert_{2}$ denotes the Euclidean norm of a vector. Furthermore, vector $\bm{a}_i(\omega_q)$ contains the sensor responses for the $i$-th \ac{PLD}, and matrix $\bm{D}_i = \left( \bm{I}_{N} \otimes [D^{0}_{\theta_i}, \, D^{1}_{\theta_i}, \, \ldots, \,  D^{R}_{\theta_i}] \right) \otimes [D^{0}_{\phi_i}, \, D^{1}_{\phi_i}, \, \ldots, \, D^{P}_{\phi_i}]$, with $\otimes$ representing the Kronecker product and $\bm{I}_{N}$ being an identity matrix of dimension $N \times N$, is of dimension $N \times N(P+1)(R+1)$ and contains all combinations of interpolation factors for the $i$-th \ac{PLD} as required by (\ref{eq:B_PFSB}). 
Moreover, the scalar $\gamma$ in the left term of (\ref{eq:OP_Constraints}) represents the lower bound on the \ac{WNG} and can be adjusted by the user. Hence, the robustness of the proposed beamformer design can be easily controlled. As in our previous work \cite{mabande:icassp2009,mabande:iwaenc2010,barfuss:hscma2017}, we use the same desired response for all frequencies. 
The optimization problem (\ref{eq:OP_LSApproximation}), (\ref{eq:OP_Constraints}) is convex \cite{mabande:iwaenc2010,mabande:phdthesis2014}. To solve it, we use CVX, a package for specifying and solving convex optimization problems \cite{cvx,grant_boyd:convexoptimization2008} in Matlab. After solving (\ref{eq:OP_LSApproximation}) and (\ref{eq:OP_Constraints}) for all $Q$ frequencies, the time-domain \ac{FIR} filters $\bm{w}_{n,p,r}$ are obtained by an \ac{FIR} approximation of the resulting optimum frequency-domain coefficients $\bm{w}_\text{f}(\omega_q)$, to ensure causality of the realized filters.%which interpolates between the $Q$ discrete frequencies and ensures causality of the realized filters.

%relation to previous work
Note that the original \ac{1D-RLSFIP} beamformer design in \cite{mabande:iwaenc2010} is obtained from the proposed \ac{2D-RLSFIP} beamformer design by setting $R=0$ and distributing the \acp{PLD} in a plane corresponding to a fixed elevation angle $\theta_i = \ang{90} \, \forall i$. 
By further setting $P=0$, the non-polynomial \ac{RLSFI} beamformer design in \cite{mabande:icassp2009} is obtained. \vspace*{-2mm}

%------------------------------------------------------------------------
%Experiments
%------------------------------------------------------------------------
\section{Evaluation}
\label{sec:evaluation}
\vspace*{-1.5mm}
In the following we present a design example of the proposed \ac{2D-RLSFIP} beamformer design and evaluate its flexible beam steering using signal-independent and signal-dependent measures.
%
%parameters of beamformer design
The beamformer design is carried out for the twelve-microphone robot head, shown in Fig.~\ref{fig:setup_headArray}, which was developed during the EU-FP7 Project EARS \cite{EARS,tourbabin:daga2016}. To account for the effects of the robot's head on the sound field, we incorporate measured \acp{HRTF} as sensor responses $g_n(\omega_q,\Omega_m)$ into the beamformer design, as, e.g., in \cite{barfuss:hscma2017}.
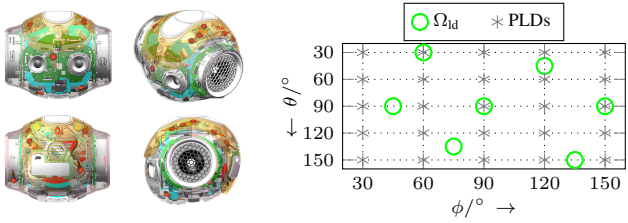
\begin{figure}[t]
	\subfigure[Microphone positions (red circles) on robot's head.]{ 
		\centering
		\raisebox{2.5mm}{\includegraphics[width = 3.25cm]{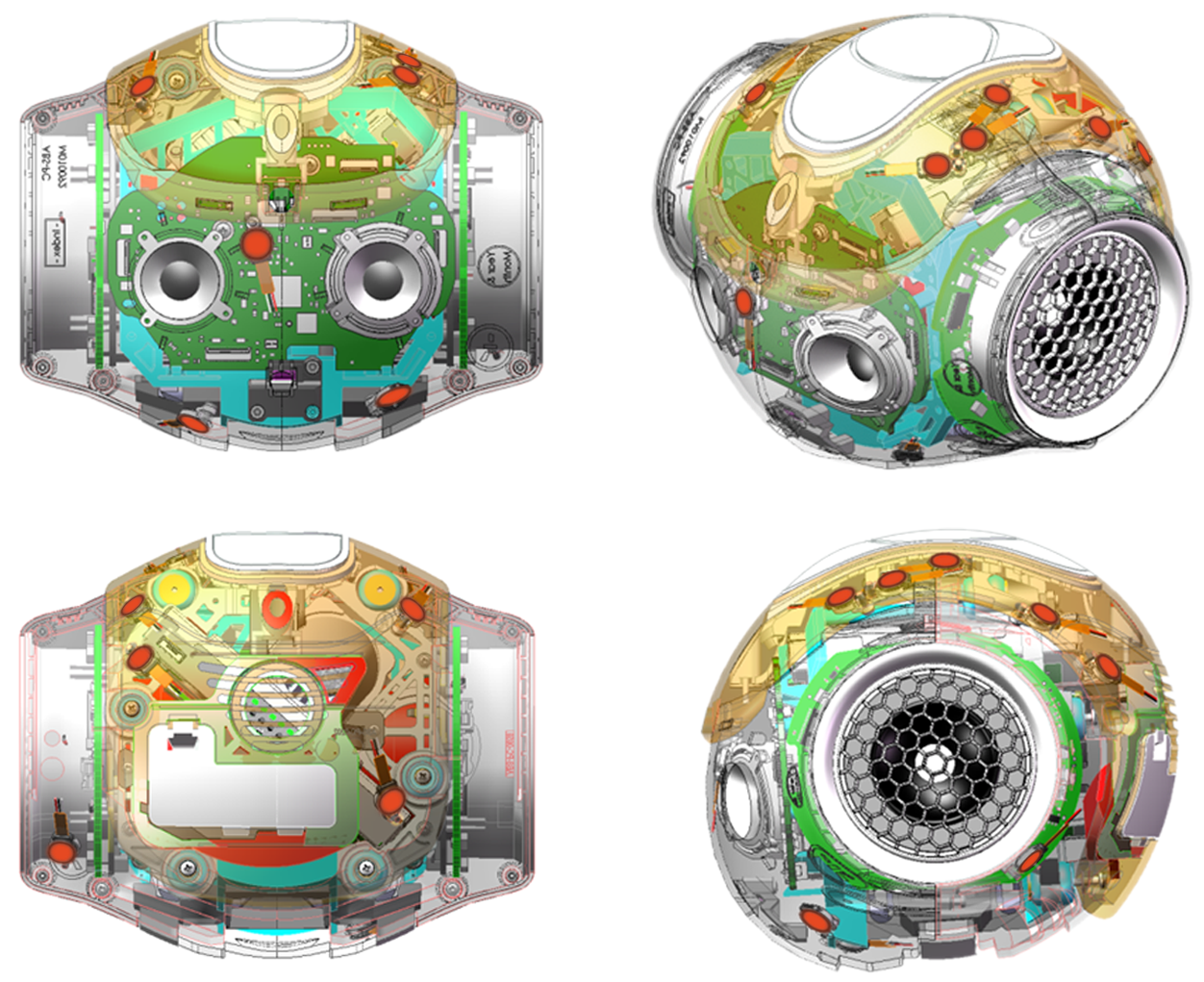}}
		\label{fig:setup_headArray}
	}\hspace{1mm}
	\subfigure[\acp{PLD} (gray asterisks) and evaluated source positions (green circles).]{ 
%		\centering  
		\begin{tikzpicture}
			\begin{axis}[
				label style = {font=\scriptsize},
				tick label style = {font=\scriptsize},
				width=0.3*\textwidth,height=3.25cm,grid=major,grid style = {dotted,black},  				
				axis on top, 	
				axis on top,
				ylabel style={yshift=-2mm}, 
				xlabel style={yshift=1mm},
				xtick={30,60,90,120,150},
				ytick={30,60,90,120,150},
				xlabel={$\phi/ \SIUnitSymbolDegree \, \rightarrow$},	  
				ylabel={$\leftarrow \, \theta/ \SIUnitSymbolDegree$},
				ymin=20,ymax=160,xmin=20,xmax=160,
				y dir=reverse,
				legend style={at={(.5,1.325)},anchor=north,legend columns=-1,font=\scriptsize,/tikz/every even column/.append style={column sep=10pt}},
				]
				%look directions
				\addplot+[only marks,mark size=3pt,mark=o,thick,green] coordinates{
					(45,90) (60,30) (75,135) (90,90) (120,45) 
					(135,150) (150,90)}; \addlegendentry{$\Omega_\text{ld}$}
				%PLDs
				\addplot+[only marks,mark size=2.5pt,mark=asterisk, gray] coordinates{
					(30,30) (60,30) (90,30) (120,30) (150,30) 
					(30,60) (60,60) (90,60) (120,60) (150,60) 
					(30,90) (60,90) (90,90) (120,90) (150,90) 
					(30,120) (60,120) (90,120) (120,120) (150,120) 
					(30,150) (60,150) (90,150) (120,150) (150,150)}; \addlegendentry{\acp{PLD}}
			\end{axis}
		\end{tikzpicture}
		\label{fig:PLDs_evaluated_look_directions}
	}  
	\vspace{-5mm}
	\caption{Illustration of (a) microphone positions (red circles) at the $12$-microphone humanoid robot's head, and (b) of the evaluated source positions  (green circles) and \acp{PLD} of the proposed \ac{2D-RLSFIP} beamformer design (gray asterisks).}
	\vspace{-5mm}
\end{figure}
%messungen von HRTFs
The time-domain \acp{HRIR} were measured in a low-reverberation chamber ($T_{60} \approx \SI{50}{\milli\second}$) for a total of $2522$ loudspeaker positions distributed on a sphere with a radius of $1.1$m around the robot's head (in discrete steps of five degrees in azimuth and elevation direction). The measurements were carried out using maximum-length sequences (see, e.g., \cite{schroeder:jasa1979}), and the measured impulse responses were truncated to exclude unwanted reflections due to additional objects in the room. Due to mechanical constraints, the \acp{HRIR} were measured without the robot's torso.
%parameters beamformer design
For the beamformer design, we used a lower bound on the \ac{WNG} of  $10\text{log}_{10}\gamma=\SI{-20}{\decibel}$ and an \ac{FIR} filter length of $L=512$ samples at a sampling rate of $f_\text{s}=\SI{16}{\kilo\hertz}$.
%Desired response
An exemplary desired response used for the design is illustrated in Fig.~\ref{fig:desired_BFResponse}. As in \cite{barfuss:hscma2017}, it is specified for $M=2522$ design look directions with steps of five degrees in both directions. Each direction is represented by a rectangle, where the actual value of $\hat{B}(\omega_q,\Omega_m)$ is color-coded. 
The desired beamformer response is equal to one for the target look direction $\Omega_\text{ld}=(\ang{115},\ang{45}$) and decreases to zero at all sides, with a $3$-\si{\decibel} beamwidth of \ang{20}. For brevity, only the frontal hemisphere is shown, since the desired beamformer response in the rear hemisphere contains only zeros.
%distribution PLDs
For the polynomial beamformer design, we distribute $I=25$ \acp{PLD} uniformly on the $\phi$-$\theta$ plane in steps of $\ang{30}$ in a range of $\ang{30} \leq \Omega_i \leq \ang{150}$, as illustrated in Fig.~\ref{fig:PLDs_evaluated_look_directions}, where each \ac{PLD} is represented by a gray asterisk. Hence, flexible beam steering with sufficient performance is only possible in this angular region, which we chose with a robot audition scenario in mind, where a target source is usually standing approximately in front of the robot. 
Note that the employed distribution of \acp{PLD} is a straightforward extension of the \ac{1D-RLSFIP} beamformer design in \cite{mabande:iwaenc2010}, and is motivated by the fact that we successively apply one-dimensional interpolation in azimuth and elevation direction. 
More elaborate \ac{PLD} distributions like uniform or nearly-uniform distributions on a sphere or hemisphere (see, e.g., \cite[Chapter~3]{rafaely:book2015}) should lead to a lower overall approximation error, but may require more sophisticated interpolation strategies, which we will consider in future work.
\begin{figure}[t]
	\centering
	\begin{tikzpicture}
	\begin{axis}[
	label style = {font=\scriptsize},
	tick label style = {font=\scriptsize},
	width=0.4\textwidth,height=0.5*0.4\textwidth,
	axis on top,
	ylabel style={yshift=-2mm}, 
	xlabel style={yshift=1mm},
	xtick={0,45,90,135,180},
	ytick={0,45,90,135,180},
	xlabel={$\phi/ \SIUnitSymbolDegree \, \rightarrow$},	  
	ylabel={$\leftarrow \, \theta/ \SIUnitSymbolDegree$},
	ymin=-2.5,ymax=182.5,xmin=-2.5,xmax=182.5,
	y dir=reverse,
	scatter/use mapped color={draw=mapped color,fill=mapped color},
	colorbar,
	colorbar style={at={(1.015,1)},yticklabel style={font=\tiny,xshift=-0.75mm},width=0.2cm,height=1.975cm},
	point meta min=0, point meta max=1,
	colormap/parula,
	]
	\addplot graphics [xmin=-2.5, xmax=182.5, ymin=-2.5, ymax=182.5] {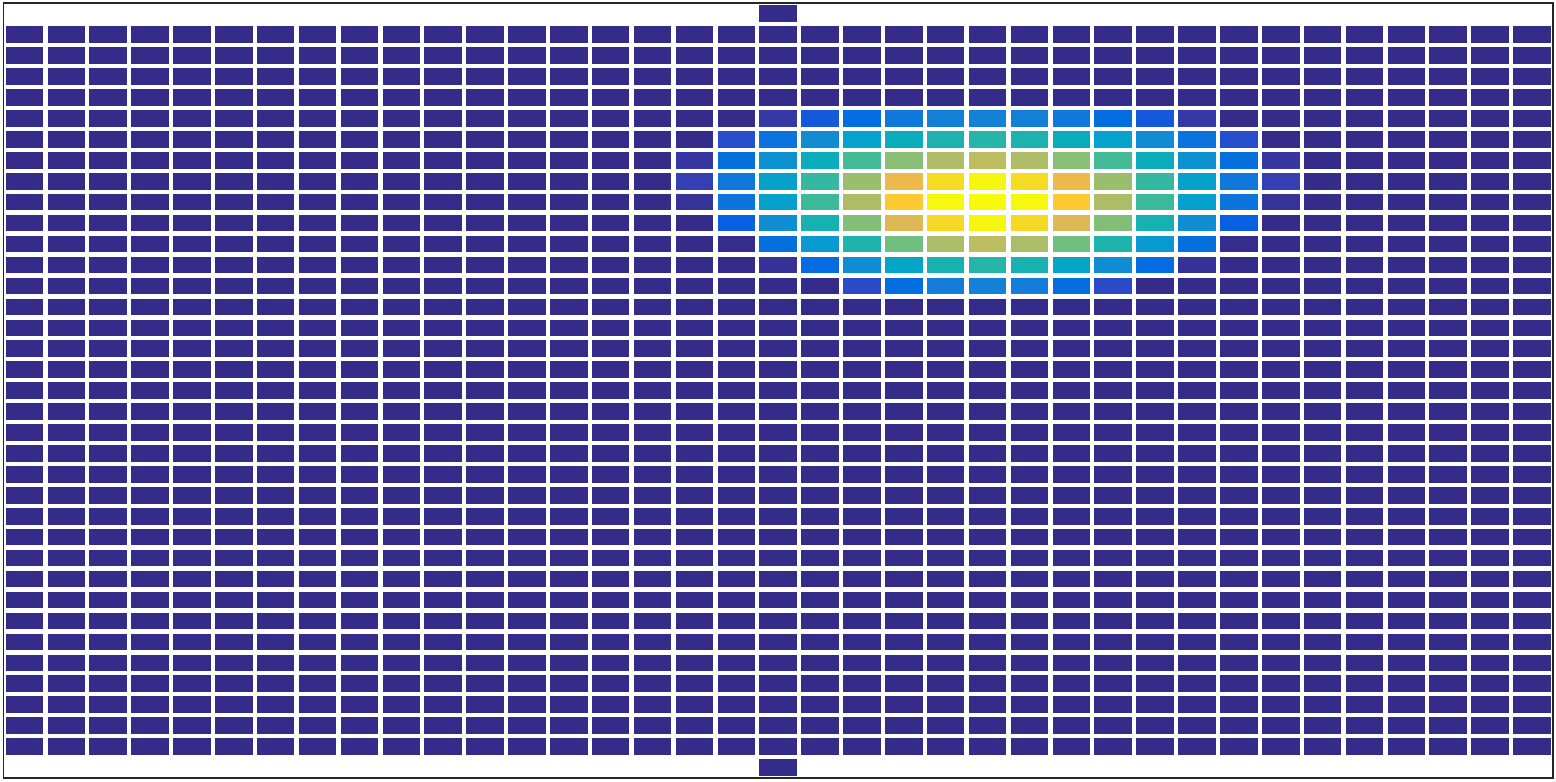};
	\end{axis}
	\end{tikzpicture}
	\vspace{-5.5mm}
	\caption{Illustration of an exemplary desired beamformer response for target look direction $\Omega_\text{ld}=(\ang{115},\ang{45})$, which is used for the \ac{RLSFI} and \ac{2D-RLSFIP} beamformer design illustrated in Fig.~\ref{fig:3D_beampattern_115_45}.}
	\label{fig:desired_BFResponse}	
	\vspace{-6mm}
\end{figure}

%Beampatterns
%------------------------------------------------------------------------
At first, we present an example to illustrate the effect of polynomial interpolation on the resulting beamformer. In Figs.~\ref{fig:3D_beampattern_115_45_RLSFI} and \ref{fig:3D_beampattern_115_45_RLSFIP} the beampatterns of the \ac{RLSFI} and \ac{2D-RLSFIP} beamformer designs for $\Omega_\text{ld}=(\ang{115},\ang{45})$ at frequency $f=\SI{2}{\kilo\hertz}$ are illustrated. The corresponding \ac{WNG} and \ac{DI} (after (2.19) in \cite{bitzer:2001superdirective}) for $\SI{300}{\hertz} \leq f \leq \SI{5}{\kilo\hertz}$ (chosen with the application to speech signal capture in mind) are given in Figs.~\ref{fig:3D_beampattern_115_45_WNG} and \ref{fig:3D_beampattern_115_45_DI}. The beampatterns were computed with \acp{HRTF} modeling the acoustic system. Thus, they effectively show the transfer function between source position and beamformer output. 
\begin{figure}[t]
	\vspace{3mm}
	\subfigure{
		\hspace{5mm}
		\begin{tikzpicture}[scale=1,trim axis left,baseline=(current axis.south)]
		\node at (-0.65,1.7) {\scriptsize (a)};
		\node at (5.8,2) {\scriptsize $20\text{log}_{10} |B(f=\SI{2}{\kilo\hertz},\Omega)| / \si{\decibel}$};   
		\begin{axis}[
		label style = {font=\scriptsize},
		tick label style = {font=\scriptsize},   
		ylabel style={yshift=-2mm}, 
		xlabel style={yshift=1mm},
		width=0.275*\textwidth,height=3.25cm,grid=major,grid style = {dotted,black},  		
		axis on top, 	
		enlargelimits=false,
		xmin=0, xmax=180, ymin=0, ymax=180,
		xtick={0,45,90,135,180},
		ytick={0,45,90,135,180},
		xlabel={$\phi/ \SIUnitSymbolDegree \, \rightarrow$},
		ylabel={$\leftarrow \, \theta/ \SIUnitSymbolDegree$},
		y dir=reverse,
		%colorbar
		colorbar horizontal, 
		colormap/parula, 		
		colorbar style={
			at={(0,1.25)}, anchor=north west, font=\tiny, width=3.85cm, height=0.2cm, xticklabel pos=upper
		},
		point meta min=-40, point meta max=0]
		\addplot graphics [xmin=-1, xmax=180, ymin=-1, ymax=180] {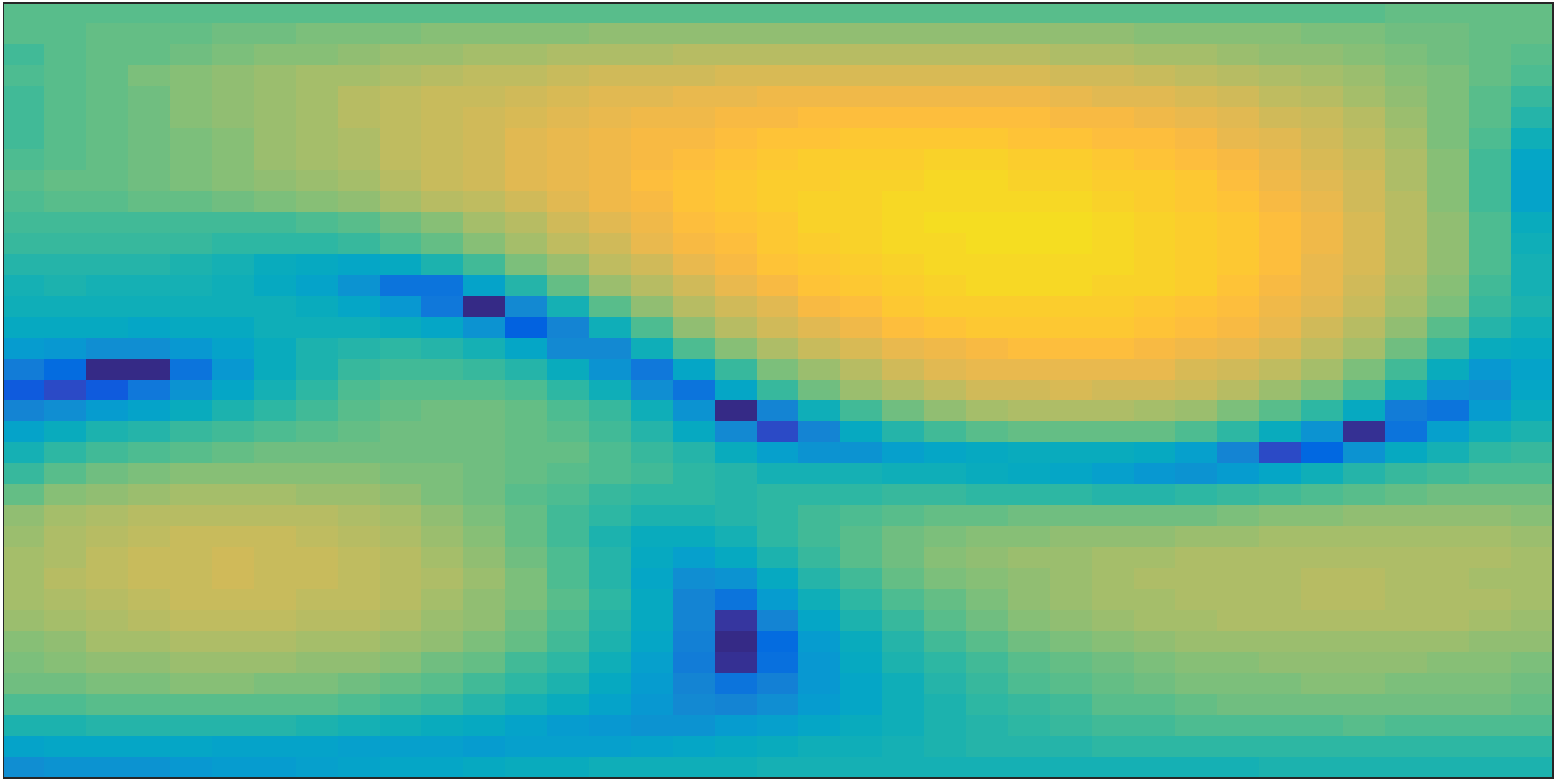};
		\end{axis}
		\end{tikzpicture}
		\label{fig:3D_beampattern_115_45_RLSFI}			
	}%\\%[-4mm]	
	\subfigure{  
		\hspace{-44.5mm}
		\begin{tikzpicture}[scale=1,baseline=(current axis.south)]
		\node at (-0.65,1.7) {\scriptsize (b)};
		\begin{axis}[
		label style = {font=\scriptsize},
		tick label style = {font=\scriptsize},   
		ylabel style={yshift=-2mm}, 
		xlabel style={yshift=1mm},
		width=0.275*\textwidth,height=3.25cm,grid=major,grid style = {dotted,black},  				
		axis on top, 	
		enlargelimits=false,
		xmin=0, xmax=180, ymin=0, ymax=180,
		xtick={0,45,90,135,180},
		ytick={0,45,90,135,180},
		xlabel={$\phi/ \SIUnitSymbolDegree \, \rightarrow$},
		ylabel={$\leftarrow \, \theta/ \SIUnitSymbolDegree$},
		y dir=reverse,		
		]
		\addplot graphics [xmin=-1, xmax=180, ymin=-1, ymax=180] {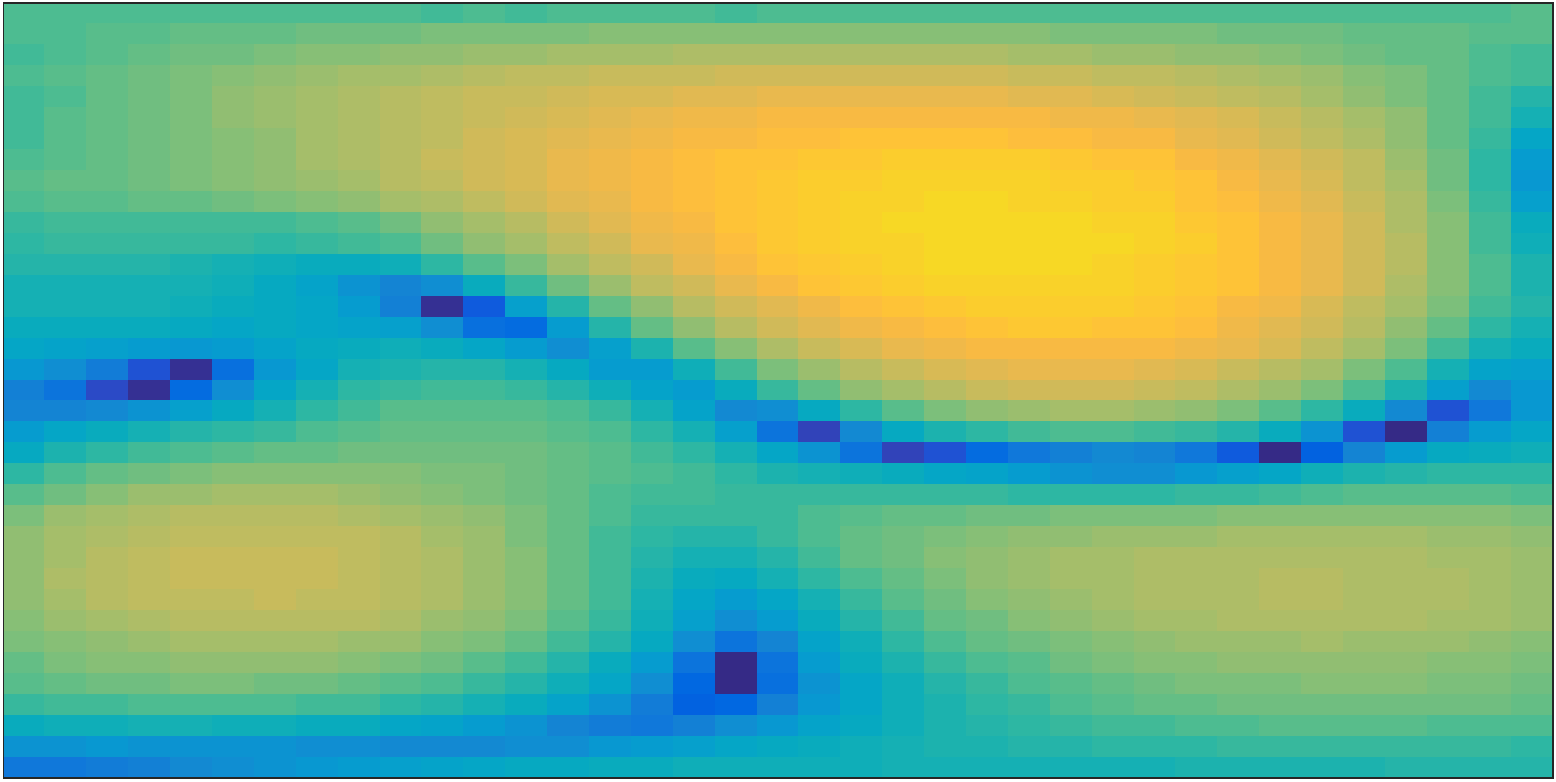};
		\end{axis}
		\end{tikzpicture}  
		\label{fig:3D_beampattern_115_45_RLSFIP}		
	}\\[-5mm]
	\subfigure{  
		\hspace{5mm}
		\begin{tikzpicture}[scale=1,trim axis left,baseline=(current axis.south)]
		\node at (-0.65,1.7) {\scriptsize (c)};
		\begin{axis}[
		label style = {font=\scriptsize},
		tick label style = {font=\scriptsize},
		ylabel style={yshift=-2mm}, 
		xlabel style={yshift=1mm},
		legend style={font=\scriptsize, yshift=0.25mm, at={(.515,0.97)}},
		width=0.275*\textwidth,height=3.25cm,grid=major,grid style = {dotted,black},
		xtick={300,1000,2000,3000,4000,5000},
		xticklabels={0.3, 1, 2, 3, 4, 5},
		xlabel={$f/ \si{\kilo\hertz} \, \rightarrow$},
		ytick={-25, -20, -15, -10, -5, 0, 5}, 
		yticklabels={{}, -20, {}, -10, {}, 0, {}}, 
		ylabel={$\text{WNG}/\si{\decibel} \, \rightarrow$},
		ymin=-25, ymax=5, xmin=300,xmax=5000]   
		\addplot[thick,blue,solid] table [x index=0, y index=1]{WNG_N_12_ArrayShape_BenchmarkII_DesResShape_2D_3dBBeamwidth_20_LookDir_az_el_115_45_WNG_-20dB_fs_16000.dat};
		\addplot[thick,myGreen,dashed] table [x index=0, y index=1]{WNG_N_12_ArrayShape_BenchmarkII_DesResShape_2D_3dBBeamwidth_20_P_4_L_4_distPLDs_uniform_numPLDs_25_LookDir_az_el_115_45_WNG_-20dB_fs_16000.dat};		
		\end{axis}       
		\end{tikzpicture}    
		\label{fig:3D_beampattern_115_45_WNG}	
	}\subfigure{  
	\hspace{-2.6mm}
	\begin{tikzpicture}[scale=1,baseline=(current axis.south)]
	\node at (-0.65,1.7) {\scriptsize (d)};
	\begin{axis}[
	label style = {font=\scriptsize},
	tick label style = {font=\scriptsize},
	ylabel style={yshift=-2mm}, 
	xlabel style={yshift=1mm},		
	legend style={font=\scriptsize, yshift=0.25mm, at={(.515,0.97)}},
	width=0.275*\textwidth,height=3.25cm,grid=major,grid style = {dotted,black},
	xtick={300,1000,2000,3000,4000,5000},
	xticklabels={0.3, 1, 2, 3, 4, 5},
	xlabel={$f/ \si{\kilo\hertz} \, \rightarrow$},
	ylabel={$\text{DI}/\si{\decibel} \, \rightarrow$},
	ymin=0, ymax=15, xmin=300,xmax=5000,
	legend pos=south east]   
	\addplot[thick,blue,solid] table [x index=0, y index=1]{DI_N_12_ArrayShape_BenchmarkII_DesResShape_2D_3dBBeamwidth_20_LookDir_az_el_115_45_WNG_-20dB_fs_16000.dat}; \addlegendentry{RLSFI};
	\addplot[thick,myGreen,dashed] table [x index=0, y index=1]{DI_N_12_ArrayShape_BenchmarkII_DesResShape_2D_3dBBeamwidth_20_P_4_L_4_distPLDs_uniform_numPLDs_25_LookDir_az_el_115_45_WNG_-20dB_fs_16000.dat}; \addlegendentry{2D-RLSFIP};
	\end{axis}       
	\end{tikzpicture}    
	\label{fig:3D_beampattern_115_45_DI}	
}	
\vspace{-6.5mm}
\caption{Illustration of beampatterns of the (a) \acs{RLSFI} and (b) \ac{2D-RLSFIP} beamformer designs at $f=\SI{2}{\kilo\hertz}$, with  $\Omega_\text{ld}=(\ang{115},\ang{45})$ and $10\text{log}_{10}\gamma=\SI{-20}{\decibel}$. \ac{WNG} and \acs{DI} are shown in sub-figures (c) and (d).}
\label{fig:3D_beampattern_115_45}
\vspace{-5.5mm}
\end{figure}
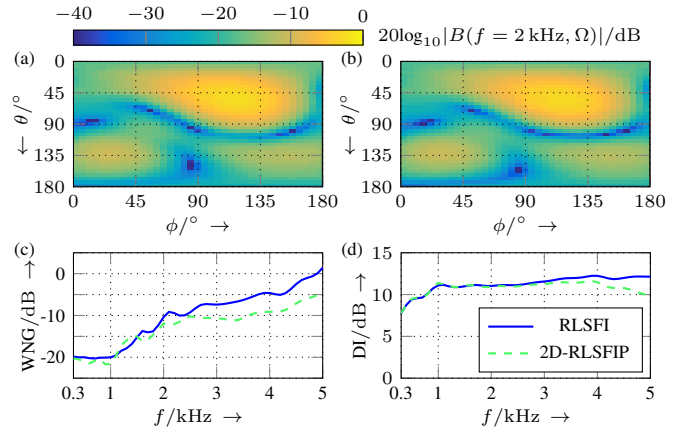
Since the target look direction lies between four \acp{PLD}, approximation errors of the \ac{RLSFI} by the \ac{2D-RLSFIP} beamformer due to polynomial interpolation are to be expected.
Nevertheless, the two beampatterns look very similar. However, when looking at Figs.~\ref{fig:3D_beampattern_115_45_WNG} and \ref{fig:3D_beampattern_115_45_DI}, it becomes obvious that the \ac{WNG} and \ac{DI} of the two beamformer designs are different:
While the \ac{WNG} of the \ac{2D-RLSFIP} beamformer is mostly lower than that of the \ac{RLSFI} beamformer over the entire frequency range, the \ac{DI} is only slightly lower for $f > \SI{3}{\kilo\hertz}$. It can also be seen that the \ac{WNG} of the polynomial beamformer is only slightly lower than the imposed constraint for $f \leq \SI{1}{\kilo\hertz}$ due to polynomial interpolation. So far, there is no direct way to ensure that the \ac{WNG} constraint is fulfilled for target look directions unequal to the \acp{PLD}. If the violation of the \ac{WNG} constraint is too severe, the \ac{PPF} order and position of the \acp{PLD} need to be adapted. 
Further investigation of the two beamformer designs have confirmed that when the target look direction is equal to one of the \acp{PLD}, the \ac{RLSFI} and {2D-RLSFIP} beamformer designs yield almost identical results. It could also be confirmed that the beamformer response of the \ac{2D-RLSFIP} beamformer is similarly flat as of the \ac{RLSFI} beamformer for frequencies below  $f \leq \SI{5}{\kilo\hertz}$.

%MSE
%------------------------------------------------------------------------
To demonstrate how accurately the proposed \ac{2D-RLSFIP} beamformer design can approximate the \ac{RLSFI} beamformer design, we now investigate the \ac{MSE} between the magnitudes of the beamformer responses of the two beamformers for all $Q$ frequencies:\vspace*{-1.5mm}
\begin{equation}
	\textstyle
%	\scriptsize
	\text{MSE}(\Omega_\text{ld}) = \sum\limits_{q=1}^{Q}\sum\limits_{m=1}^{M} \frac{ \left( \left\vert B^\text{RLSFI}_{D_{\phi_\text{ld}},D_{\theta_\text{ld}}}(\omega_q,\Omega_m) \right\vert - \left\vert B^\text{2D-RLSFIP}_{D_{\phi_\text{ld}},D_{\theta_\text{ld}}}(\omega_q,\Omega_m) \right\vert \right)^{2}}{Q \cdot M},
	\label{eq:MSE}
		\vspace*{-0.5mm}
\end{equation}
where $B^\text{RLSFI}_{D_{\phi_\text{ld}},D_{\theta_\text{ld}}}(.)$ and $B^\text{2D-RLSFIP}_{D_{\phi_\text{ld}},D_{\theta_\text{ld}}}(.)$ denote the beamformer response of the \ac{RLSFI} and \ac{2D-RLSFIP} beamformer design steered to $\Omega_\text{ld}$. In addition, we calculate the mean \ac{DI}, $\overline{\text{DI}}_\text{RLSFI}$ and $\overline{\text{DI}}_\text{2D-RLSFIP}$, of each beamformer over all frequencies, and take the difference thereof: $\Delta \overline{\text{DI}}(\Omega_\text{ld}) = \overline{\text{DI}}_\text{RLSFI}(\Omega_\text{ld})-\overline{\text{DI}}_\text{2D-RLSFIP}(\Omega_\text{ld})$. Both, \ac{MSE} as well as $\Delta \overline{\text{DI}}$ should be as small as possible. In the following, \ac{MSE} and $\Delta \overline{\text{DI}}$ are evaluated for the entire angular range of interest $\ang{30} \leq \Omega_\text{ld} \leq \ang{150}$, and for the entire frequency range. %, and for a frequency range of $\SI{300}{\hertz} \leq f \leq \SI{5}{\kHz}$. 
The results are illustrated in Figs.~\ref{fig:MSE} (\ac{MSE}) and \ref{fig:DI_diff} ($\Delta \overline{\text{DI}}$), respectively.
%MSE
Both \ac{MSE} and $\overline{\text{DI}}$ are relatively small over most of the angular range of interest, with values equal to zero at the \acp{PLD}. In between the \acp{PLD}, slightly larger values occur due to polynomial interpolation, with a maximum \ac{MSE} of $\text{MSE}_\mathrm{max}=0.04$ and a maximum $\Delta \overline{\text{DI}}$ of $\Delta \overline{\text{DI}}_\text{max}=\SI{2.8}{\decibel}$. Note that if we only evaluate the frequency range up to $\SI{5}{\kilo\hertz}$, \ac{MSE} and $\Delta\overline{\text{DI}}$ are reduced to a great extent, with maximum values of $\text{MSE}_\mathrm{max}=0.01$ and $\Delta \overline{\text{DI}}_\text{max}=\SI{1.2}{\decibel}$. This shows that in the frequency range which is most relevant for speech signal capture, polynomial approximation works very accurately, and that most of the approximation error can be found in the higher frequency range above $\SI{5}{\kilo\hertz}$. The slight asymmetries in Figs.~\ref{fig:MSE} and \ref{fig:DI_diff} can be attributed to the asymmetric sensor placement on the robot's head (cf. Fig.~\ref{fig:setup_headArray}).
\begin{figure}[t]
	\hspace{-4mm}
	\subfigure[$\text{MSE}(\Omega_\text{ld})$ in (\ref{eq:MSE}).]{
		\begin{tikzpicture}
		\begin{axis}[
		label style = {font=\scriptsize},
		tick label style = {font=\scriptsize},   
		ylabel style={yshift=-2.5mm}, 
		xlabel style={yshift=1mm},
		width=0.27*\textwidth,height=3.25cm,grid=major,grid style = {dotted,black},  		
		axis on top, 	
		enlargelimits=false,
		xmin=30, xmax=150, ymin=30, ymax=150,
		xtick={30,60,90,120,150},
		ytick={30,60,90,120,150},
		xlabel={$\phi_\text{ld}/ \SIUnitSymbolDegree \, \rightarrow$},
		ylabel={$\leftarrow \, \theta_\text{ld}/ \SIUnitSymbolDegree$},
		y dir=reverse,
		%colorbar
		colorbar,
		colorbar horizontal,
		colorbar style={at={(0,1.15)},yticklabel style={font=\tiny,xshift=-0.75mm,scaled ticks=false},height=0.2cm,xtick={0,0.01,0.02,0.03,0.04,0.05},xticklabels={0,0.01,0.02,0.03,0.04,0.05},xticklabel pos=upper},
		point meta min=0, point meta max=0.05,
		colormap/parula]
		\addplot graphics [xmin=30, xmax=150, ymin=30, ymax=150] {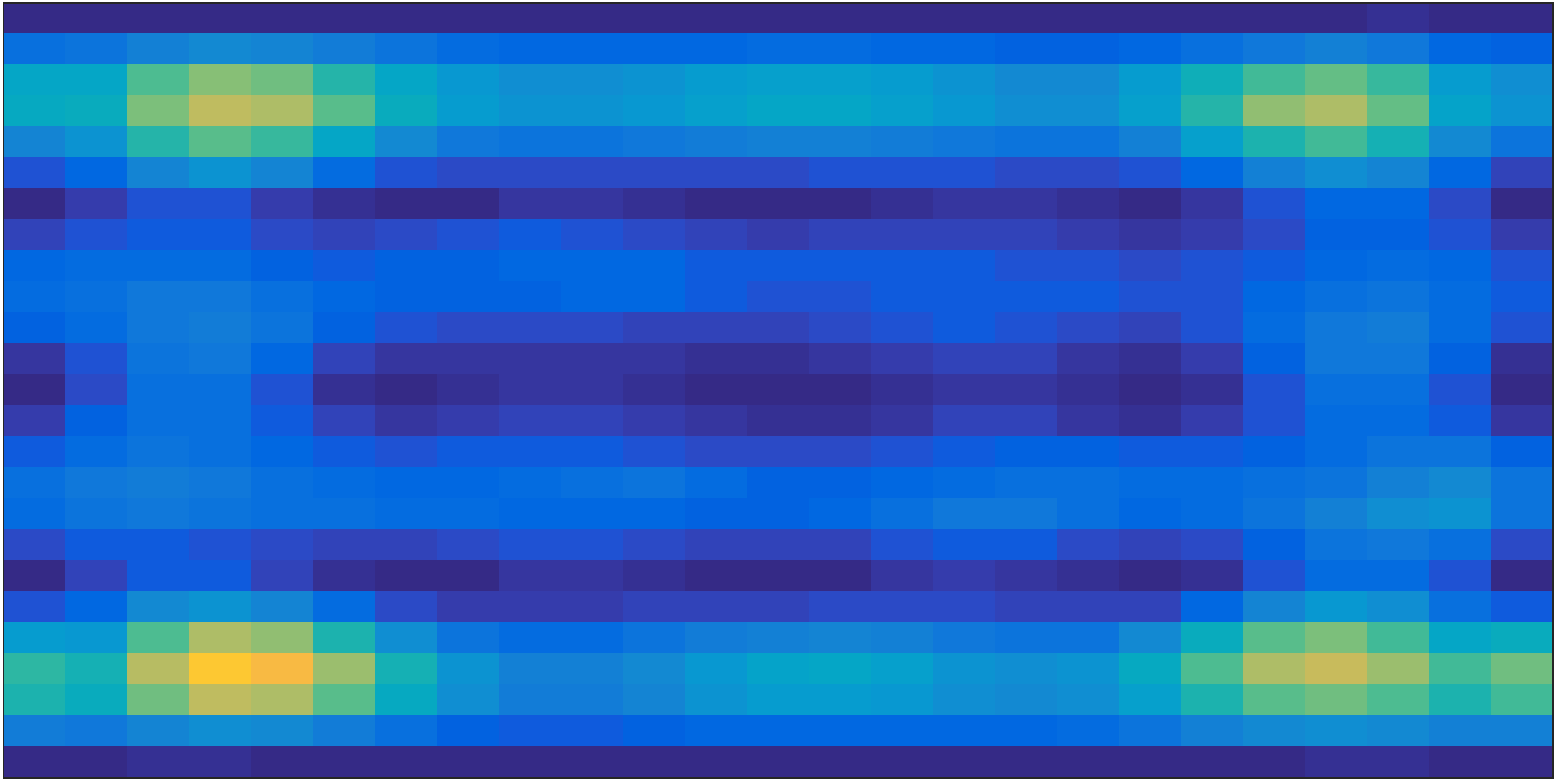};
		\end{axis}
		\end{tikzpicture}
		\label{fig:MSE}	
		\hspace{-2mm}		
	}%
	\subfigure[Difference $\Delta \overline{\text{DI}}(\Omega_\text{ld})$ of mean \acp{DI}.]{
		\begin{tikzpicture}
				\node at (3,1.825) {\scriptsize $\si{\decibel}$};   
		\begin{axis}[
		label style = {font=\scriptsize},
		tick label style = {font=\scriptsize},   
		ylabel style={yshift=-2.5mm}, 
		xlabel style={yshift=1mm},
		width=0.27*\textwidth,height=3.25cm,grid=major,grid style = {dotted,black},  		
		axis on top, 	
		enlargelimits=false,
		xmin=30, xmax=150, ymin=30, ymax=150,
		xtick={30,60,90,120,150},
		ytick={30,60,90,120,150},
		xlabel={$\phi_\text{ld}/ \SIUnitSymbolDegree \, \rightarrow$},
		ylabel={$\leftarrow \, \theta_\text{ld}/ \SIUnitSymbolDegree$},
		y dir=reverse,
		%colorbar
		colorbar,
		colorbar horizontal,
		colorbar style={at={(0,1.15)},yticklabel style={font=\tiny,xshift=-0.75mm},xticklabel pos=upper,width=2.75cm,height=0.2cm},
		point meta min=0, point meta max=3,
		colormap/parula]
		\addplot graphics [xmin=30, xmax=150, ymin=30, ymax=150] {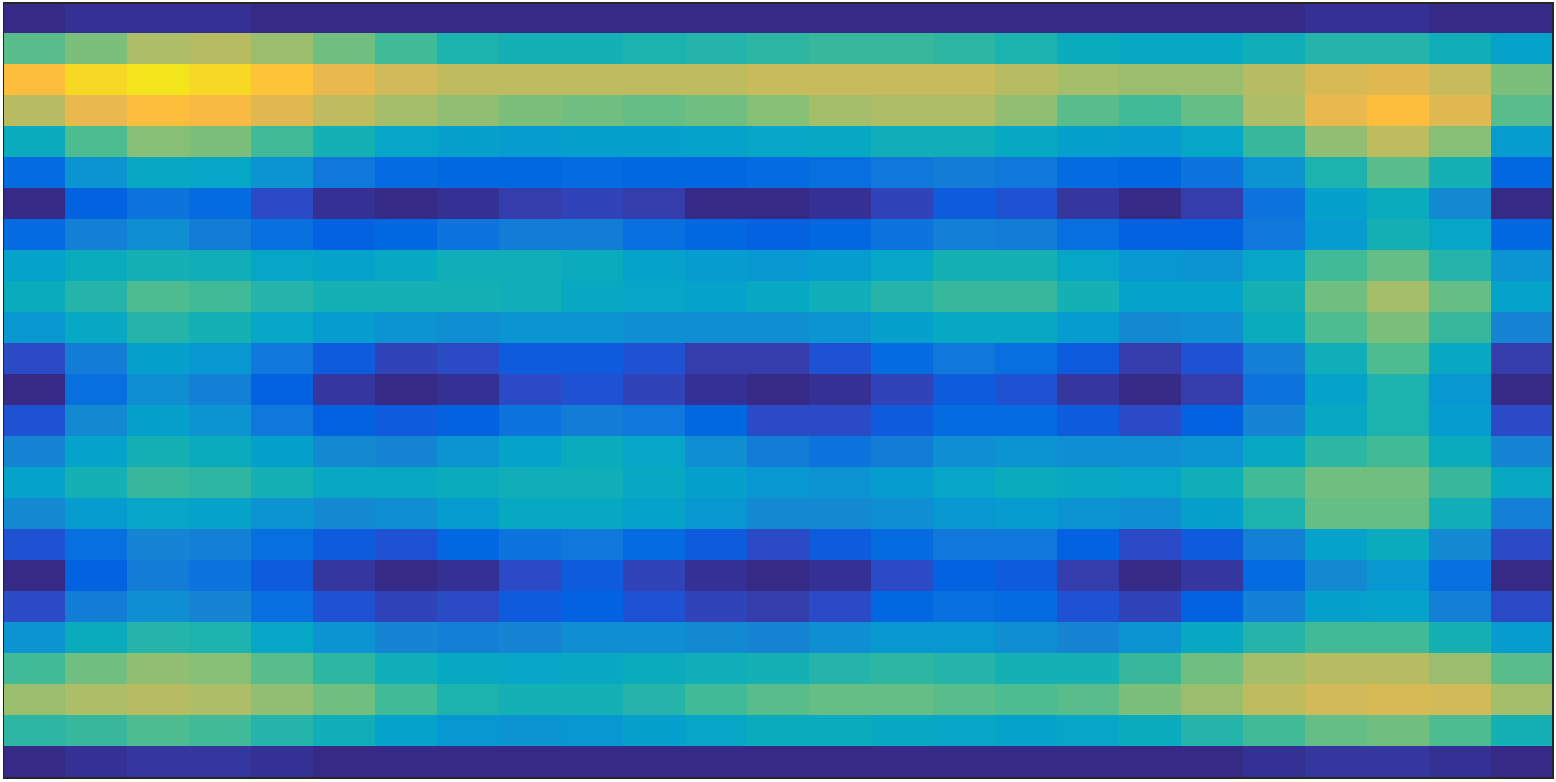};
		\end{axis}
		\end{tikzpicture}
		\label{fig:DI_diff}			
	}
		\vspace{-5mm}
		\caption{Illustration of (a) \ac{MSE} between beamformer responses of \acs{RLSFI} and \ac{2D-RLSFIP} beamformers, and difference $\Delta \overline{\text{DI}}(\Omega_\text{ld})$ of mean \acp{DI} for $\ang{30} \leq \Omega_\text{ld} \leq \ang{150}$.}
		\label{fig:MSE_DI}
		\vspace{-5mm}
\end{figure}

%fwSegSnr
%------------------------------------------------------------------------
Finally, we evaluate the signal enhancement performance of the \ac{RLSFI} and \ac{2D-RLSFIP} beamformer designs in a two-speaker scenario. As performance measure, we use the \ac{fwSegSNR} \cite{hu:tasl2008}, where we select the desired signal components at the frontmost microphone and at the beamformer's output as reference signal for calculating the input and output \ac{fwSegSNR}.
%scenario
The evaluated source positions, illustrated in Fig.~\ref{fig:PLDs_evaluated_look_directions} as green circles, were chosen such that some of them coincide with \acp{PLD} of the polynomial beamformer design, and some do not.
Each target source position is evaluated six times with an interfering speaker located at one of the remaining six positions. The \ac{fwSegSNR} was calculated for each combination of target and interfering source position and averaged over the six different \ac{fwSegSNR} values. The resulting average target source position-specific \ac{fwSegSNR} levels are summarized in Fig.~\ref{fig:results_fwSegSNR}.
%
%acoustic environment and measurement of HRTFs and RIRs
The microphone signals were created by convolving clean speech signals of duration $\SI{20}{\second}$ with \acp{RIR}, which were measured in the same low-reverberation chamber as the \acp{HRIR}. In addition, white Gaussian noise was added to each microphone channel with a \ac{SNR} of $\SI{40}{\decibel}$ to model sensor noise. 
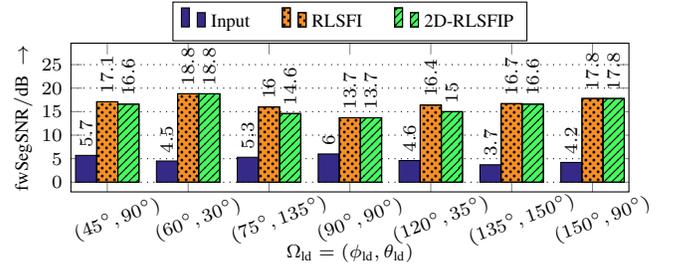
\begin{figure}[t]
	\hspace{7.5mm}
	\begin{tikzpicture}[scale=1,trim axis left]
		\begin{axis}[
		width=9cm,height=0.5*0.4\textwidth,
		ymajorgrids=true,
		grid style = {dotted,black},
		label style = {font=\scriptsize},
		tick label style = {font=\scriptsize},
		xlabel style={yshift=2mm},
		ylabel style={yshift=-1mm},
		ybar=0pt, % space of 0pt between adjacent bars
		bar width=8pt,
		enlargelimits=0.075,
		ytick={0,5,10,15,20, 25},
		xtick={0,30,60,90,120,150,180},
		xtick align=inside,
		xticklabels={{$(\ang{45},\ang{90})$}, {$(\ang{60},\ang{30})$}, {$(\ang{75},\ang{135})$}, {$(\ang{90},\ang{90})$}, {$(\ang{120},\ang{35})$}, {$(\ang{135},\ang{150})$}, {$(\ang{150},\ang{90})$}},		
		x tick label style={rotate=20,anchor=north,yshift=-1.5mm},
		xlabel shift={-1mm},
		xlabel={$\Omega_\text{ld}=(\phi_\text{ld},\theta_\text{ld})$},
		ylabel={$\text{fwSegSNR}/\text{dB}\, \rightarrow$},
		xmin=0, xmax=180, ymin=0, ymax=27.5,
		every node near coord/.append style={font=\scriptsize, rotate=90, anchor=west, /tikz/.cd},
		nodes near coords, nodes near coords align={vertical},
		legend style={at={(0.5,1.28)},anchor=north,legend columns=-1,font=\scriptsize,/tikz/every even column/.append style={column sep=10pt}},
		legend entries={Input, RLSFI, 2D-RLSFIP},
		]		
		%SNR=infinity dB
%		%FWSEGSNR_in  5.7	4.6	5.4	6.1	4.7	3.7 4.3
%		\addplot[black,fill=blue, postaction={pattern=north east lines}] coordinates {(0,5.7) (30,4.6) (60,5.4) (90,6.1) (120,4.7) (150,3.7) (180,4.3)}; 
%		%FWSEGSNR_RLSFI,out	19.9	21.5	19.8	15.7	18.7	19.6	20.7
%		\addplot[black,fill=red, postaction={pattern=crosshatch dots}] coordinates {(0,19.9) (30,21.5) (60,19.8) (90,15.7) (120,18.7) (150,19.6) (180,20.7)}; 
%		%FWSEGSNR_RLSFIP,out	19.5	21.5	18.8	15.7	17.7	19.5	20.7
%		\addplot[black,fill=green] coordinates {(0,19.5) (30,21.5) (60,18.8) (90,15.7) (120,17.7) (150,19.5) (180,20.7)}; 
		%SNR=40dB
		%FWSEGSNR_in  5.7	4.5	5.3	6.0	4.6	3.7	4.2
		\addplot[black,fill=myBlue] coordinates {(0,5.7) (30,4.5) (60,5.3) (90,6) (120,4.6) (150,3.7) (180,4.2)}; 
		%FWSEGSNR_RLSFI,out	17.1	18.8	16.0	13.7	16.4	16.7	17.8
		\addplot[black,fill=myOrange, postaction={pattern=crosshatch dots}] coordinates {(0,17.1) (30,18.8) (60,16) (90,13.7) (120,16.4) (150,16.7) (180,17.8)}; 		
		%FWSEGSNR_RLSFIP,out	16.6	18.8	14.6	13.7	15.0	16.6	17.8
		\addplot[black,fill=myGreen,postaction={pattern=north east lines}] coordinates {(0,16.6) (30,18.8) (60,14.6) (90,13.7) (120,15) (150,16.6) (180,17.8)}; 
		\end{axis}
	\end{tikzpicture}	
	\vspace{-8mm}
	\caption{Average target source position-specific \acp{fwSegSNR} in \si{\decibel}, obtained at the input (blue bars) and output of the \ac{RLSFI} (orange bars) and \ac{2D-RLSFIP} (green bars) beamformers.}
	\label{fig:results_fwSegSNR}
	\vspace{-5.5mm}
\end{figure}
The results confirm our previous observations: When the target look direction $\Omega_\text{ld}$ coincides with one of the \acp{PLD}, the results of the \ac{2D-RLSFIP} and \ac{RLSFI} beamformers are identical. When the target source is not located in one of the \acp{PLD}, the \ac{fwSegSNR} levels of the polynomial beamformer are slightly lower than, but still very close to, those of the \ac{RLSFI} beamformer. In all cases a significant enhancement of the target source can be observed. 
A brief comparison of the signal enhancement performance of the \ac{1D-RLSFIP} and \ac{2D-RLSFIP} beamformers showed that the latter outperforms the former when the target look direction lies directly between the \acp{PLD} as for $(\Omega_\text{ld})=(\ang{75},\ang{135})$.
\vspace*{-4.5mm}

%------------------------------------------------------------------------
%Conclusion
%------------------------------------------------------------------------
\section{conclusion}
\label{sec:conclusion}
\vspace*{-2mm}
%What we did
In this work, we proposed a design method for a robust two-dimensional polynomial beamformer, formulated as a convex optimization problem, which allows for flexible beam steering in both azimuth and elevation direction. The beamformer's robustness can be easily controlled by the user by adjusting a single scalar value for each angular dimension in the optimization problem.
The proposed polynomial beamformer design method was applied to a twelve-element microphone array, integrated into the head of a humanoid robot, and was evaluated using signal-independent and signal-dependent measures in a robot audition scenario. The results confirmed the efficacy of the polynomial beamformer design, i.e., the \ac{2D-RLSFIP} beamformer approximates the \ac{RLSFI} very accurately. As a consequence, the \ac{2D-RLSFIP} beamformer is an effective method for robust and flexible time-domain real-time beamforming.
%future work: evaluation of P and L, different distributions of PLDs, different interpolation approaches
Future work includes analyzing the influence of \ac{PPF} orders on the beamforming performance, the distribution of \acp{PLD} in the desired angular range, and alternative interpolation approaches.%, and an evaluation in more reverberant acoustic environments.

%------------------------------------------------------------------------
%References
%------------------------------------------------------------------------
\bibliographystyle{IEEEtran}
\bibliography{barfuss_waspaa_2017}

\end{sloppy}
\end{document}